\documentclass[conference]{IEEEtran}
\usepackage[left=0.70in, right=0.70in, top=1in, bottom=1in]{geometry}
\IEEEoverridecommandlockouts
\usepackage[pdftex]{graphicx}
\usepackage[ruled]{algorithm2e}
\usepackage{algorithmic}
\usepackage[small]{caption}
\usepackage{float}
\usepackage{xspace}
\usepackage{fancyhdr}
\usepackage{amssymb,amsmath}
\usepackage{subfig}
\usepackage{xcolor,amsmath}
\usepackage{epstopdf}
\usepackage{bbding}
\usepackage{cite} 
\usepackage{color}
\begin{document}

\title{\huge{Reconfigurable Intelligent Surfaces Enhanced NOMA D2D Communications Underlaying UAV Networks}}
\author{Wali Ullah Khan$^\dagger$,  Eva Lagunas$^\dagger$, Asad Mahmood$^\dagger$, Zain Ali$^\ddagger$, Symeon Chatzinotas$^\dagger$, Bj\"orn Ottersten$^\dagger$ \\$^\dagger$Interdisciplinary Centre for Security, Reliability and Trust (SnT), University of Luxembourg\\
$^\ddagger$Department of Electrical and Computer Engineering, University of California, Santa Cruz, USA\\
\{waliullah.khan, eva.lagunas, asad.mahmood, symeon.chatzinotas, bjorn.ottersten\}@uni.lu\\
zainalihanan1@gmail.com
%\thanks{This work was supported by the Luxembourg National Research Fund via
%project 5G-Sky, ref. FNR/C19/IS/13713801/5G-Sky.}
}%

%\markboth{Submitted to IEEE}%

\maketitle

\begin{abstract}
%Device-to-device (D2D) communication offers high spectral efficiency, low energy consumption, and short transmission with low latency. Due to limited spectrum resources, D2D communication share spectrum with the cellular system as an underlay, which can cause co-channel interference and degrade the quality-of-service (QoS) for cellular users. Effective management of interference and optimization of spectrum sharing between D2D communication and the cellular system is necessary to ensure desired QoS for both D2D and cellular users. 
Device-to-device (D2D) communications offers high spectral efficiency, low energy consumption and transmission latency. However, one of the main limitations of D2D communications is co-channel interference from underlaying wireless system. Reconfigurable intelligent surfaces (RIS) is a promising technology because it can manipulate the electromagnetic waves in their environment to overcome interference and enhance wireless communications. 
%By adjusting the phase and amplitude of the reflected waves from the RIS, it is possible to create a customized wireless communication environment that can improve signal quality, extend coverage, increase energy efficiency, and overcome interference issue. 
%It can potentially overcome the limitations of interference in D2D communication. 
This paper considers RIS enhanced D2D communications underlaying unmanned aerial vehicle (UAV) networks with non-orthogonal multiple access (NOMA). The objective is to maximize the sum rate of NOMA D2D communications by optimizing the power budget of D2D transmitter, NOMA power allocation coefficients of D2D receivers and passive beamforming of RIS while guaranteeing the quality of services of UAV user. Due to non-convexity, the optimization problem is intractable and challenging to handle. Therefore, it is solved in two parts using alternating optimization. Simulation results unviel the performance of the proposed RIS enhanced D2D communications scheme. Results demonstrate that the proposed scheme achieves 15\% and 27\% higher sum rates compared to the fixed power D2D and orthogonal D2D schemes.
\end{abstract}


\IEEEpeerreviewmaketitle

\section{Introduction}
\IEEEPARstart{T}{he rise} of mobile devices and smart terminals has resulted in an unprecedented increase in demand for wireless data traffic \cite{de2021survey}. To satisfy this demand and provide seamless communications, device-to-device (D2D) technology has emerged as a promising solution. By enabling users who are physically near each other to communicate directly, D2D communications can bypass the cellular base station (BS) and significantly reduce energy consumption while enhancing the quality of service (QoS) and offering early emergency warnings \cite{jameel2018survey}. Additionally, D2D links can share the same spectrum resources as cellular links, thus alleviating the issue of spectrum scarcity. However, interference is inevitable in D2D communications and must be managed to protect cellular networks from harmful interference and maintain communications quality \cite{yang2020spectral}. Moreover, conventional communications utilize the orthogonal multiple access (OMA) techniques for spectrum sharing, which suffers from a significant drawback - each spectrum resource can only accommodate a single user, thus, restricting the number of users that can connect to the system \cite{khan2019joint}. 

In response to the above challenges, recent advances in communications have introduced innovative technologies such as reconfigurable intelligent surfaces (RIS) and non-orthogonal multiple access (NOMA) \cite{khan2022opportunities,maraqa2020survey}. These cutting-edge solutions have received significant attention due to their ability to address the issue of energy and spectral efficiency. In particular, RIS can be employed to eliminate D2D interference and meet high data rate demands by controlling the reflection and scattering of electromagnetic waves that impinge on the surface \cite{liu2021reconfigurable}. Similarly, NOMA can achieve high spectral efficiency by giving multi-user access to the network over the same system resources \cite{elbayoumi2020noma}. 

Due to high mobility and flexibility, unmanned aerial vehicles (UAVs) can be rapidly deployed and moved to various locations, making them highly flexible communication platforms \cite{9915455}. This agility allows them to serve as temporary base stations, communication relays, or aerial hot-spots in areas with limited infrastructure. Moreover, UAVs can be dispatched to provide instant connectivity in emergencies or events where temporary high-capacity communication is needed, such as concerts, festivals, or disaster response scenarios. UAVs can be efficiently integrated into existing wireless communications networks and used to optimize system performance by filling coverage gaps, enhancing signal strength in crowded areas, or reducing interference \cite{10008627}.

The combination of D2D communications with NOMA and UAVs has shown great potential for enhancing wireless networks but also poses significant challenges as this area of research is still in its early stages. One major challenge is the power control for NOMA users, as optimal power allocation is necessary to guarantee reliable signal decoding at the receiver side. Another challenge arises is the interference from the underlaying UAV communications. In this situation, RIS can provide reflective paths to enhance signal quality and minimize interference between cellular and D2D communications. Therefore, optimising the phase shifts of the RIS is critical to ensure the received signal quality is improved and interference is minimized \cite{10133841}. Another crucial factor is power control for NOMA users, as optimal power allocation is necessary to guarantee reliable signal decoding at the receiver side. 

Recently, some research works have integrated RIS with D2D communications. In \cite{yang2021reconfigurable}, Yang {\em et al.} have investigated the spectrum and energy efficiency of RIS enhanced D2D communications through joint optimization of power and beamforming at RIS and BS. The authors of \cite{ji2022reconfigurable} have maximized the sum rate of D2D system by optimizing the phase shift and position of RIS and power of D2D. The research work in \cite{yang2022sum} has optimized the transmit power of D2D communications and beamforming of RIS to improve the average D2D rate. Moreover, Jia {\em et al.} \cite{jia2020reconfigurable} have optimized the energy consumption of D2D communications by controlling the transmit power and beamforming at RIS. Cai {\em et al.} \cite{cai2020reconfigurable} have maximize the ergodic weighted sum rate of RIS enhanced D2D communications through efficient resource management of the system. The paper in \cite{cao2021sum} has optimized user pairing, transmit power and BS beamforming to enhance the sum rate of RIS-assited D2D communications. Furthermore, Peng {\em et al.} \cite{peng2021ris} have optimized the phase shift of RIS to improve the achievable rate of the system. In addition, some authors have also studied the effective capacity \cite{shah2022statistical}, outage probability \cite{nguyen2022performance}, and physical layer security of RIS enhanced D2D communications \cite{khoshafa2020reconfigurable}. 
%%%%%%%%%%%%%%%%
\begin{figure}[t]
\centering
\includegraphics [width=0.45\textwidth]{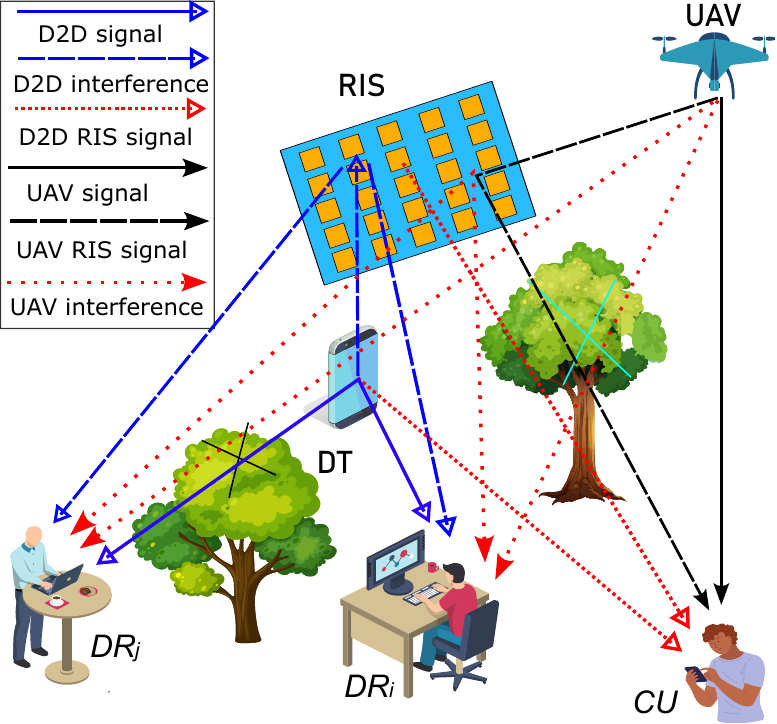}
\caption{System model}
\label{SM1}
\end{figure}
%%%%%%%%%%%%%

Based on the existing literature, RIS is combined with D2D communications using orthogonal spectrum resources. To the best of the our knowledge, the problem of RIS enhanced NOMA D2D communications underlaying UAV network has not yet been investigated. Therefore, this work considers a new framework of RIS enhanced NOMA D2D communications underlaying UAV network which is open topic to study. The objective is to maximize the sum rate of the system by optimizing the power budget of D2D transmitter, power allocation coefficients of NOMA D2D receivers and passive beamforming of RIS while ensuring the quality of services for UAV network. The remaining work can be organized as follows. Section II discusses system model and problem formulation of RIS enhanced NOMA D2D communications underlaying UAV network. Section III provided the solution of sum rate maximization problem. Section IV presents numerical results and Section V concludes this work.

\section{System Model and Problem Formulation}
We consider a RIS enhanced NOMA D2D communications underlaying cellular UAV networks, which consists of one UAV acting as an aerial BS, one cellular user, a D2D group and a RIS system, as illustrated in Fig. \ref{SM1}. The cellular and D2D transmissions share the same spectrum simultaneously. Therefore, for both cellular and D2D communications, the cellular receiver and D2D receivers get signals from both UAV and D2D transmitters. In the D2D group, a D2D transmitter communicates with two D2D receivers using the NOMA protocol. The RIS system is strategically mounted to enhance the communications of both cellular and D2D communications. Resultantly, cellular and D2D receivers can receive direct signals from their transmitters and reflected ones through the RIS system. It is assumed that both cellular and D2D communications follow single antenna configuration while the channel state information is available in the whole network. Let us denote the cellular user as $CU$ while D2D receivers as $DR_i$ and $DR_j$, respectively. We consider that RIS consists of $K\times K$ elements such that $\boldsymbol\Phi=\text{diag}\{\varphi_1,\varphi_2,\varphi_3,\dots, \varphi_K\}$, where $\varphi_k$ denotes the reconfigurable coefficent of element $k$. Moreover, $\varphi_k$ should satisfy $|\varphi_k|=1, \ \forall k\in K$. 

Next, we study the transmitted and received signals of cellular and D2D communications. If $s$ represents the transmit signal of UAV for $CU$ and $x$ denotes the superimposed signal of DT for $DR_i$ and $DR_j$. Then, the transmit signal of UAV and DT can be defined as $s=\sqrt{q_c}s$ and $x=\sqrt{P_t\Lambda_i}x_i+\sqrt{P_t\Lambda_j}x_j$, where $q_c$ is the transmit power of $CU$ and $P_t$ denotes the transmit power of DT. In addition, $\Lambda_i$ and $\Lambda_j$ are the power allocation coefficient of $DR_i$ and $DR_j$. Based on the transmit signals, the received signals at $CU$, $DR_i$ and $DR_j$ can be expressed as\footnote{Following \cite{10122731}, we consider general probabilistic channels between different terminals, we omit here the details for simplicity and space.}:
\begin{align}
y_c=(g_{t_1,c}+g_{t_1,k}\Phi g_{k,c})s+(h_{t_2,c}+h_{t_2,k}\Phi h_{k,c})x+n_c,\label{1}
\end{align}
\begin{align}
y_i=(h_{t_2,i}+h_{t_2,k}\Phi h_{k,i})x+(g_{t_1,i}+g_{t_1,k}\Phi g_{k,i})s+n_i,\label{2}
\end{align}
\begin{align}
y_j=(h_{t_2,j}+h_{t_2,k}\Phi h_{k,j})x+(g_{t_1,j}+g_{t_1,k}\Phi g_{k,j})s+n_j,\label{3}
\end{align}
where in (\ref{1}), $g_{t_1,c}$ is the direct channel gain from UAV to $CU$, $g_{t_1,k}$ denotes the channel gain from UAV to RIS, $g_{k,c}$ represents the channel gain from RIS to $CU$, $h_{t_2,c}$ stats the direct interference channel gain from DT to $CU$, $h_{t_2,k}$ is the interference channel gain from DT to RIS, $h_{k,c}$ denotes the interference channel gain from RIS to $CU$, and $n_c$ is the additive white Gaussian noise (AWGN) with zero mean and $\sigma^2$ variance. Similarly, in (\ref{2}), $h_{t_2,i}$ represents the direct channel gain from DT to $DR_i$, $h_{t_2,k}$ is the channel gain from DT to RIS, $h_{k,i}$ denotes the channel gain from RIS to $DR_i$, $g_{t_1,i}$ is the direct interference channel gain from UAV to $DR_i$, $g_{t_1,k}$ is the interference channel gain from UAV to RIS, $g_{k,i}$ stats the interference channel gain from RIS to $DR_i$, and $n_i$ is the AWGN with zero mean and $\sigma^2$ variance. Accordingly, in (\ref{3}), $h_{t_2,j}$ is the direct channel gain from DT to $DR_j$, $h_{t_2,k}$ is the channel gain from DT to RIS, $h_{k,j}$ is the channel gain from RIS to $DR_j$, $g_{t_1,j}$ denotes the direct interference channel gain from UAV to $DR_j$, $g_{t_1,k}$ represents the channel gain from UAV to RIS, $g_{k,j}$ is the interference channel gain from RIS to $DR_j$, and $n_j$ shows AWGN with zero mean and $\sigma^2$ variance. 

Following the NOMA protocol, D2D receiver with strong channel conditions can apply SIC to remove the interference from D2D receiver having weak channel conditions. Based on our system design, we assume that $DR_i$ has stronger channel conditions than $DR_j$. Then, the rates of $CU$, $DR_i$ and $DR_j$ can be derived as $R_c=\log_2(1+\gamma_c)$, $R_i=\log_2(1+\gamma_i)$ and $R_j=\log_2(1+\gamma_j)$, where $\gamma_c$, $\gamma_i$ and $\gamma_j$ are the signals to interference plus noise ratios (SINRs) and can be defined as:
\begin{align}
\gamma_c=\frac{q_c|g_{t_1,c}+g_{t_1,k}\Phi g_{k,c}|^2}{P_t(\Lambda_i+\Lambda_j)|h_{t_2,c}+h_{t_2,k}\Phi h_{k,c}|^2+\sigma^2},\label{4}
\end{align}
\begin{align}
\gamma_i=\frac{P_t\Lambda_i|h_{t_2,i}+h_{t_2,k}\Phi h_{k,i}|^2}{q_c|g_{t_1,i}+g_{t_1,k}\Phi g_{k,i}|^2+\sigma^2}, \label{5}
\end{align}
\begin{align}
\gamma_j=\frac{P_t\Lambda_i|h_{t_2,j}+h_{t_2,k}\Phi h_{k,j}|^2}{\Omega_{i,j}+q_c|g_{t_1,j}+g_{t_1,k}\Phi g_{k,j}|^2+\sigma^2}, \label{6}
\end{align}
where in (\ref{4}), the denominator shows the interference of D2D communications plus noise variance. In (\ref{5}), the denominator denotes the interference of cellular communications plus noise variance. In (\ref{6}), $\Omega_{i,j}=P_t\Lambda_i|h_{t_2,j}+h_{t_2,k}\Phi h_{k,j}|^2$ is the NOMA interference from $DR_i$ and the second term is the interference from cellular communications.

We aims to enhance the achievable sum rate of RIS enhanced NOMA D2D communications while ensuring the QoS of cellular communications. This can be achieved by simultaneously optimizing the power budget of DT, power allocation of $DR_i$ and $DR_j$ along with passive beamforming of RIS. The maximization problem of sum capacity can be mathematically formulated as follows:
%%%%%%%%%%%%%%%%%%%%%%%%%%%
\begin{alignat}{2}
&\underset{(P_t,\boldsymbol{\Lambda}, \boldsymbol{\Phi})}{\text{maximize}} \ (R_i+R_j)\label{OP} \\
 s.t.\ &    \gamma_{c}\geq \gamma_{min}, \tag{7a}\label{7a}\\
& 0\leq P_t(\Lambda_{i}+\Lambda_{j})\leq P_{max},\tag{7b} \label{7b}\\
& |\varphi_k|=1,\ \forall k\in K,\tag{7c} \label{7c}\\
& \Lambda_{i}+\Lambda_{j}\leq 1,\tag{7d} \label{7d}\\
 & 0\leq\Lambda_{i}\leq1, 0\leq\Lambda_{j}\leq1,\tag{7e} \label{7e}
\end{alignat}
%%%%%%%%%%%%%%%%%%%%%%%%%%
where (\ref{OP}) is the objective function of sum rate maximization. Constrain (\ref{7a}) ensures the QoS cellular user while (\ref{7b}) limits the transmit power of DT. Constraint (\ref{7c}) adds for passive beamforming at RIS. Constraints (\ref{7d}) and (\ref{7e}) control the power allocation according to the NOMA protocol.

\section{Proposed Sum Rate Maximization Scheme}
We can observe that the formulated problem in (\ref{OP}) is non-convex due to the interference in the rate expressions. Moreover, it is also coupled on two optimization variables, i.e., transmit power at DT and passive beamforming at RIS. As a result, the problem is intractable and obtaining a joint optimal solution is very complex and challenging. Therefore, we adopt alternating optimization in which the sum rate maximization problem in (\ref{OP}) is solved in two parts. First, we calculate the power budget of DT and power allocation for $DR_i$ and $DR_j$ with any given beamforming at RIS. Then, in the second part, given the power budget $P_t$ and power allocation of $\Lambda_i$ and $\Lambda_j$, we calculate the passive beamforming at RIS.
\subsection{Power Allocation at DT}
Given the phase shift at RIS, the power allocation problem can be re-expressed as follows:
%%%%%%%%%%%%%%%%%%%%%%%%%%%
\begin{alignat}{2}
&\underset{(P_t,\Lambda_i,\Lambda_j)}{\text{maximize}} \ (R_i+R_j)\label{OP1} \\
 s.t.\ &  (7a),(7b),(7d).\tag{8a}\label{8a}
\end{alignat}
%%%%%%%%%%%%%%%%%%%%%%%%%%
This problem is still non-convex due to the rate expressions. To further reduce the complexity, we exploit SCA method. Using this method, the rate expressions of $DR_i$ and $DR_j$ can be re-written as $\bar{R}_i=\alpha_i\log_2(\gamma_i)+\beta_i$ and $\bar{R}_j=\alpha_j\log_2(\gamma_j)+\beta_j$, where $\alpha_\nu$ and $\beta_\nu$ can be stated as:
\begin{align}
\alpha_\nu=\frac{\gamma_\nu}{1+\gamma_\nu},\ \nu\in\{i,j\},
\end{align}
\begin{align}
\beta_\nu=\log_2(1+\gamma_\nu)-\alpha_\nu\log_2(\gamma_\nu),\ \nu\in\{i,j\}.
\end{align}
The problem in (\ref{OP1}) can be efficiently updated as follows:
%%%%%%%%%%%%%%%%%%%%%%%%%%%
\begin{alignat}{2}
&\underset{(P_t,\Lambda_i,\Lambda_j)}{\text{maximize}} \ (\bar{R}_i+\bar{R}_j)\label{OP12} \\
 s.t.\ &   q_c|g_{t_1,c}+g_{t_1,k}\Phi g_{k,c}|^2\geq \gamma_{min}\nonumber\\&\times(P_t(\Lambda_i+\Lambda_j)|h_{t_2,c}+h_{t_2,k}\Phi h_{k,c}|^2+\sigma^2),\tag{11a}\label{11a} \\
 & (7b),(7d). \tag{11b}\label{11b}
\end{alignat}
%%%%%%%%%%%%%%%%%%%%%%%%%%
Next, we employ the Lagrangian method to solve problem (\ref{OP12}), where the Lagrangian function can be derived as:
\begin{align}
&L(P_t,\Lambda_i,\Lambda_j,\boldsymbol{\eta})=(\Bar{R}_i+\Bar{R}_j)\eta_1(q_c|g_{t_1,c}+g_{t_1,k}\Phi g_{k,c}|^2\nonumber\\&-\gamma_{min}(P_t(\Lambda_i+\Lambda_j)|h_{t_2,c}+h_{t_2,k}\Phi h_{k,c}|^2+\sigma^2))\nonumber\\&+\eta_2(P_{max}-P_t)+\eta_3(1-\Lambda_i), \label{Lag}
\end{align}
where $\boldsymbol{\eta}=\{1,2,3\}$. Now we use KKT conditions such as:
\begin{align}
\frac{\partial L(P_t,\Lambda_{i},\Lambda_{j},\boldsymbol{\eta})}{\partial P_t, \Lambda_{i},\Lambda_{j}}|_{P_t,\Lambda_{i},\Lambda_{j}=P^*_t\Lambda_{i}^*,\Lambda_{j}^*}=0,
\end{align}
To calculate the power budget at DT, we calculate derivation of (\ref{Lag}) with respect to $P_t$ such as:
\begin{align}
&\frac{\partial L(.)}{\partial P_t}= P_t^2 (H_j \Lambda_i (\eta_2 + H_c \eta_1 \gamma_{min}))+P_t (-\alpha_iH_j \Lambda_i +\nonumber\\&  (\eta_2 + H_c \eta_1 \gamma_{min}) (G_j + \sigma^2))-(\alpha_i + \alpha_j) (G_j + \sigma^2),
\end{align}
After some straightforward calculations, the value of $P_t$ can be expressed as:
\begin{align}
P^*_t= \dfrac{-B \pm \sqrt{B^2 -4 A C}}{2 A},
\end{align}
where
\begin{align}
& A=H_j \Lambda_i (\eta_2 + H_c \eta_1 \gamma_{min})\\
& B=-\alpha_i H_j \Lambda_i + (\eta_2 + H_c \eta_1 \gamma_{min}) (G_j + \sigma^2) \\
& C=-(\alpha_i + \alpha_j) (G_j + \sigma^2)   
\end{align}
Next, we derive (\ref{Lag}) with respect to $\Lambda_i$, it can be written as:
\begin{align}
&\frac{\partial L(.)}{\partial \Lambda_i}=-\Lambda_i^3 H_j^2 \eta_3 P_t^2+
\Lambda_i^2 H_j P_t(\alpha_i H_j P_t -\eta_3 (G_j -H_j \nonumber\\&  P_t + \sigma^2))+\Lambda_i H_j P_t (\alpha_i (G_j - H_j P_t + \sigma^2) +\alpha_j (G_j+ H_j \nonumber\\&  P_t + \sigma^2) + \eta_3 (G_j + \sigma^2) )+H_j P_t (-\alpha_i G_j - \alpha_i \sigma^2),
\end{align}
where $H_i=|h_{t_2,i}+h_{t_2,k}\Phi h_{k,i}|^2$, $H_j=|h_{t_2,j}+h_{t_2,k}\Phi h_{k,j}|^2$, $H_c=|h_{t_2,c}+h_{t_2,k}\Phi h_{k,c}|^2$ and $G_j=q_c|g_{t_1,j}+g_{t_1,k}\Phi g_{k,j}|^2$. Note that (19) is a third-order polynomial function which can be efficiently solved by available solvers in MATLAB.

Finally, the value of $\Lambda_j$ can be efficiently calculated as:
\begin{align}
\Lambda^*_j= 1-\Lambda^*_i.
\end{align}

  %%%%%%%%%%%%%%%%%%%%%%%%%%%%%%%%%%%%%%%%%%
%   \begin{algorithm}[!t]
%    {\bf Step 1: Parameter initialization}  \\ Initialize the sets of $\mathcal M$, $\mathcal N$, the $P_m$ of each SAP, the minimum achievable capacity $C_{min}$, the values of weighted coefficient $\ell$, the minimum power gap between IoVs power $\omega$, the step size $\alpha$, the gradient $\zeta L$, and the iteration number as $t=0$.\\
%    {\bf Step 2: Parameter optimization} \\
%    \While{not converge}{Compute the correction vector $\psi$ using Equation (\ref{25})\;
%   Update the transmit power of IoVs at it serving SAP using Equation (\ref{24})\;
%     \eIf{the power of IoVs Converges}{
%     Stop the simulation\;
%     }{
%      Derive the $C$ as Equation (\ref{23}) until the power convergence\;
%      Set $t=t+1$, and go to step 2\;
%     }
%    }
%    {\bf Step 3: Programing termination}
%    \caption{Proposed alternating optimization}
%   \end{algorithm}
%%%%%%%%%%%%%%%%%%%%%%%%%%%%%%%%%%%%%%%%%%%%%%%%%%%% 
\subsection{Passive beamforming at RIS}
After computing $P^*_t$, $\Lambda_i$ and $\Lambda_j$, the optimization problem in (\ref{OP}) can be 
 further simplified. It can be noticed that the direct links from DT to $DR_i$ and $DR_j$ have no impact on the RIS beamforming. Therefore, without loss of generality, the SINRs in (\ref{4})-(\ref{6}) can be efficiently updated as:
 \begin{align}
\hat{\gamma}_c=\frac{q_c|g_{t_1,k}\Phi g_{k,c}|^2}{P_t(\Lambda_i+\Lambda_j)|h_{t_2,k}\Phi h_{k,c}|^2+\sigma^2},\label{20}
\end{align}
\begin{align}
\hat{\gamma}_i=\frac{P_t\Lambda_i|h_{t_2,k}\Phi h_{k,i}|^2}{q_c|g_{t_1,k}\Phi g_{k,i}|^2+\sigma^2}, \label{21}
\end{align}
\begin{align}
\hat{\gamma}_j=\frac{P_t\Lambda_i|h_{t_2,k}\Phi h_{k,j}|^2}{P_t\Lambda_i|h_{t_2,k}\Phi h_{k,j}|^2+q_c|g_{t_1,k}\Phi g_{k,j}|^2+\sigma^2}, \label{22}
\end{align}
 where (\ref{20}), (\ref{21}) and (\ref{22}) are now the SINRs of $CU$, $DR_i$ and $DR_j$ from RIS. As a result, the problem in (\ref{OP}) can be reformulated for passive beamforming at RIS as follows:
%%%%%%%%%%%%%%%%%%%%%%%%%%%
\begin{alignat}{2}
&\underset{(\boldsymbol{\Phi})}{\text{maximize}} \ (\log_2(1+\hat{\gamma_i})+\log_2(1+\hat{\gamma}_j))\label{OP2} \\
 s.t.\ &  \hat{\gamma}_c\geq \gamma_{min},\ (\ref{7c}),\tag{24a}\label{22a}
\end{alignat}
%%%%%%%%%%%%%%%%%%%%%%%%%%
Next, we define a diagonal vector of RIS elements such as $\boldsymbol{\psi}=[\psi_1,\psi_2,\psi_3,\dots,\psi_K]$, where $\boldsymbol{\psi}\in\mathbb C^{K\times1}$ is a re-arrange vector of $\boldsymbol{\Phi}$ and $\psi_k=\varphi^H_k$. Now we introduce auxiliary vectors such as $\boldsymbol{\bar{g}}_{c}=g_{t_1,k}\circ g_{k,c}$, $\boldsymbol{\bar{h}}_{i}=h_{t_2,k}\circ h_{k,i}$, and $\boldsymbol{\bar{h}}_{j}=h_{t_2,k}\circ h_{k,j}$, where $\circ$ stats Hadamard product. Moreover, the following equality $|g_{t_1,k}\Phi g_{k,c}|^2=|\boldsymbol{\psi}^H\boldsymbol{\bar{h}}_c|^2$, $|h_{t_2,k}\Phi h_{k,i}|^2=|\boldsymbol{\psi}^H\boldsymbol{\bar{h}}_i|^2$ and $|h_{t_2,k}\Phi h_{k,j}|^2=|\boldsymbol{\psi}^H\boldsymbol{\bar{h}}_j|^2$ hold, which can be easily verified. By incorporating these changes, the problem in (\ref{OP2}) can be re-expressed as follows:
%%%%%%%%%%%%%%%%%%%%%%%%%%%%%%%%%%%%%%%%%%
\begin{figure*}[!t]
\begin{align}
&\log_2\bigg(1+\frac{\text{Tr}(\boldsymbol{\Psi}\boldsymbol{H}_{i})}{\text{Tr}(\boldsymbol{\Psi}\boldsymbol{G}_c)+\sigma^2}\bigg)+\log_2\bigg(1+\frac{\text{Tr}(\boldsymbol{\Psi}\boldsymbol{H}_{j})}{\text{Tr}(\boldsymbol{\Psi}\boldsymbol{\bar{H}}_{j}+\boldsymbol{\Omega}\boldsymbol{\Bar{G}}_j)+\sigma^2}\bigg)=(\log_2(\text{Tr}(\boldsymbol{\Psi}\boldsymbol{G}_c)+\sigma^2+\text{Tr}(\boldsymbol{\Psi}\boldsymbol{H}_{i}))\nonumber\\
&-\log_2(\text{Tr}(\boldsymbol{\Psi}\boldsymbol{G}_c)+\sigma^2))+(\log_2(\text{Tr}(\boldsymbol{\Psi}\boldsymbol{\bar{H}}_{j}+\boldsymbol{\Omega}\boldsymbol{\Bar{G}}_j)+\sigma^2+\text{Tr}(\boldsymbol{\Psi}\boldsymbol{H}_{j}))-\log_2(\text{Tr}(\boldsymbol{\Psi}\boldsymbol{\bar{H}}_{j}+\boldsymbol{\Omega}\boldsymbol{\Bar{G}}_j)+\sigma^2)). \tag{32}\label{253}
\end{align}\hrulefill
\end{figure*}
%%%%%%%%%%%%%%%%%%%%%%%%%%%%%%%%%%%%%%%%%%
 %%%%%%%%%%%%%%%%%%%%%%%%%%%
\begin{alignat}{2}
&\underset{(\boldsymbol{\psi})}{\text{maximize}}  \ \log_2\bigg(1+\frac{P_t\Lambda_i|\boldsymbol{\psi}^H\boldsymbol{\bar{h}}_i|^2}{q_c|\boldsymbol{\psi}^H\boldsymbol{\bar{g}}_c|^2+\sigma^2}\bigg)\label{OP21}\\&+\log_2\bigg(1+\frac{P_t\Lambda_j|\boldsymbol{\psi}^H\boldsymbol{\bar{h}}_j|^2}{P_t\Lambda_i|\boldsymbol{\psi}^H\boldsymbol{\bar{h'}}_j|^2+q_c|\boldsymbol{\psi}^H\boldsymbol{\bar{g'}}_j|^2+\sigma^2}\bigg)\nonumber \\
 s.t.\  &   q_c|\boldsymbol{\psi}^H\boldsymbol{\bar{g}}_c|^2\geq \bar\gamma_{min}(P_t(\Lambda_i+\Lambda_j)|\boldsymbol{\psi}^H\boldsymbol{\bar{h'}}_c|^2+\sigma^2), \tag{25a}\label{23a}\\
   & |\psi_k|=1, k\in K,\tag{25b}\label{23b}
\end{alignat}
%%%%%%%%%%%%%%%%%%%%%%%%%%
where $\boldsymbol{\bar{h'}}_j=h_{t_2,k}\circ h_{k,j}$, $\boldsymbol{\bar{g'}}_j=g_{t_1,k}\circ g_{k,j}$ and $\boldsymbol{\bar{h'}}_c=h_{t_2,k}\circ h_{k,c}$.
To further explore the optimization problem (\ref{OP21}), $P_t\Lambda_i|\boldsymbol{\psi}^H\boldsymbol{\bar{h}}_i|^2$ and $P_t\Lambda_j|\boldsymbol{\psi}^H\boldsymbol{\bar{h}}_j|^2$ can be re-expressed as:
\begin{align}
P_t\Lambda_i|\boldsymbol{\psi}^H\boldsymbol{\bar{h}}_i|^2=\boldsymbol{\psi}^HP_t\Lambda_i\boldsymbol{\bar{h}}_i\boldsymbol{\bar{h}}^H_i\boldsymbol{\psi},
\end{align}
\begin{align}
\text{Tr}(\boldsymbol{\psi}^HP_t\Lambda_i\boldsymbol{\bar{h}}_i\boldsymbol{\bar{h}}^H_i\boldsymbol{\psi})=\text{Tr}(P_t\Lambda_i\boldsymbol{\bar{h}}_i\boldsymbol{\bar{h}}^H_i\boldsymbol{\psi}\boldsymbol{\psi}^H)
\end{align}
\begin{align}
P_t\Lambda_j|\boldsymbol{\psi}^H\boldsymbol{\bar{h}}_j|^2=\boldsymbol{\psi}^HP_t\Lambda_j\boldsymbol{\bar{h}}_j\boldsymbol{\bar{h}}^H_j\boldsymbol{\psi},
\end{align}
\begin{align}
\text{Tr}(\boldsymbol{\psi}^HP_t\Lambda_j\boldsymbol{\bar{h}}_j\boldsymbol{\bar{h}}^H_j\boldsymbol{\psi})=\text{Tr}(P_t\Lambda_j\boldsymbol{\bar{h}}_j\boldsymbol{\bar{h}}^H_j\boldsymbol{\psi}\boldsymbol{\psi}^H)
\end{align}
Now we introduce new auxiliary matrices $\boldsymbol{H}_i=P_t\Lambda_i\boldsymbol{\bar{h}}_i\boldsymbol{\bar{h}}^H_i$, $\boldsymbol{H}_j=P_t\Lambda_j\boldsymbol{\bar{h}}_j\boldsymbol{\bar{h}}^H_j$, and $\boldsymbol{\Psi}=\boldsymbol{\psi}\boldsymbol{\psi}^H$. We can observe that the above matrices are positive and semi-definite and can be easily verified. By making these changes, the problem in (\ref{OP21}) can be re-formulated as:
 %%%%%%%%%%%%%%%%%%%%%%%%%%%
\begin{alignat}{2}
&\underset{(\boldsymbol{\psi})}{\text{maximize}}  \ \log_2\bigg(1+\frac{\text{Tr}(\boldsymbol{\Psi}\boldsymbol{H}_i)}{\text{Tr}(\boldsymbol{\Psi}\boldsymbol{G}_c)+\sigma^2}\bigg)\label{OP22}\\&+\log_2\bigg(1+\frac{\text{Tr}(\boldsymbol{\Psi}\boldsymbol{H}_j)}{\text{Tr}(\boldsymbol{\Psi}\boldsymbol{\hat{H}}_j+\boldsymbol{\Psi}\boldsymbol{\hat{G}}_j)+\sigma^2}\bigg)\nonumber \\
 s.t.\  &   \text{Tr}(\boldsymbol{\Psi}\boldsymbol{G}_c)\geq \bar\gamma_{min}(\text{Tr}(\boldsymbol{\Psi}\boldsymbol{\hat{H}}_c)+\sigma^2), \tag{30a}\label{28a}\\
 & \text{diag}\{\boldsymbol{\Psi}\}=1, \tag{30b}\label{28b}\\
 & \boldsymbol{\Psi}\succeq 0, \tag{30c}\label{28c}\\
 & \text{rank}(\boldsymbol{\Psi})=1, \tag{30d}\label{28d}
\end{alignat}
%%%%%%%%%%%%%%%%%%%%%%%%%%
where $\boldsymbol{\hat{H}}_j=P_t\Lambda_i\boldsymbol{\bar{h'}}_j\boldsymbol{\bar{h'}}^H_j$, $\boldsymbol{\hat{G}}_c=q_c\boldsymbol{\bar{g'}}_j\boldsymbol{\bar{g'}}^H_j$ and $\boldsymbol{\hat{H}}_c=P_t(\Lambda_i+\Lambda_j)\boldsymbol{\bar{h'}}_c\boldsymbol{\bar{h'}}^H_c$. Moreover, $\text{diag}\{\boldsymbol{\Psi}\}$ represents the diagonal elements of $\boldsymbol{\Psi}$. Therefore, $\text{diag}\{\boldsymbol{\Psi}\}=1$ similar to $\psi_k^2=1$, hence $|\psi_k|=1, \forall k\in K$. 

The optimization problem (\ref{OP22}) is still non-convex because of the objective and rank 1 constraint. In the following, we investigate the non-convexity of the problem and transform it into a convex problem. We start with the rank 1 constraint, which can be efficiently replaced by a convex semi-definite constraint as $\boldsymbol{\Psi}-\boldsymbol{\bar{\psi}}\boldsymbol{\bar{\psi}}^H\succeq 0$, where $\boldsymbol{\bar{\psi}}$ is the auxiliary variable such that $\boldsymbol{\bar{\psi}}^{K\times1}$. Now this convex semi-definite constraint can be replaced by its convex Schur complement, such as:
\begin{align}
\begin{bmatrix}
\boldsymbol{\Psi} & \boldsymbol{\bar{\psi}}\\
\boldsymbol{\bar{\psi}}^H & 1
\end{bmatrix} \succeq 0.\tag{31}
\end{align}
Next we handle the objective of (\ref{OP22}) such that it can be effectively rewritten as (\ref{253}) on the top of the page. It is a DC function and can be solved through DC programming. However, it is worth noting here that $\boldsymbol{\Psi}$ consists of both imaginary and real values. Hence, calculating its partial derivative by adopting the traditional method is very hard and challenging. Therefore, it is required to compute partial derivatives for their imaginary and real values. To do so, the problem (\ref{OP22}) transforms into semi-definite programming. Now the updated problem is also convex and can be easily solved by CVX.  
%\begin{table*}
%\centering
%\caption{Values of (21) and (22)}
%\label{tabby}
%\begin{tabular}{|c|c|} 
%\hline 
%$D_{n,m}$ & $|g_{n,m}|^2\sum\limits_{l=1}^{n-1}p_{l,m}+\sigma^2+p_{n,m}|g_{n,m}|^2+\varPhi_{n,m}+|g^{m'}_{n,m}|^2\varTheta_{n,m}$ \\\hline
%$E_{l,m}$ & $2p_{l,m}|g_{l,m}|^4(2(|g_{l,m}|^2\sum\limits_{r=1}^{l-1}p_{r,m}+g^{m'}_{l,m}\varTheta_{l,m})(2|g_{l,m}|^2)+2p_{l,m}|g_{l,m}|^4+\varPhi_{l,m})$ \\\hline
%$F_{l',m'}$ & $p_{l',m'}|g_{l',m'}|^4(2(|g_{l',m'}|^2\sum\limits_{r'=1}^{l'-1}p_{r',m'}+|g^{m''}_{r',m'}|^2\varTheta_{r',m'})(|g_{r',m'}|^2)+p_{r',m'}|g_{r',m'}|^2+\varPhi_{l',m'})$ \\\hline
%$G_{k,m}$ & $\Big(|g_{k,m}|^2\sum\limits_{u=1}^{k-1}p_{u,m}+\varPhi_{k,m}+|g^{m'}_{k,m}|^2\varTheta_{k,m}\Big)^2\Big(|g_{k,m}|^2\sum\limits_{u=1}^{k-1}p_{u,m}+\varPhi_{k,m}+|g^{m'}_{k,m}|^2\varTheta_{k,m}\Big)p_{k,m}|g_{k,m}|^2$  \\\hline
%$H_{k',m'}$ & $\Big(|g_{k',m'}|^2\sum\limits_{u'=1}^{k'-1}p_{u',m'}+\varPhi_{k',m'}+|g^{m''}_{u',m'}|^2\varTheta_{k',m'}\Big)^2+\Big(|g_{k',m'}|^2\sum\limits_{u'=1}^{k'-1}p_{u',m'}+\varPhi_{k',m'}+|g^{m''}_{k',m'}|^2\varTheta_{k',m'}\Big) p_{k',m}|g_{k',m'}|^2$ \\\hline
%\end{tabular}
%\end{table*}
%%%%%%%%%%%%%%%%%%%%%%%%%%%%%%%%%%%%%%%%%%

%%%%%%%%%%%%%%%%
\begin{figure}[!t]
\centering
\includegraphics [width=0.48\textwidth]{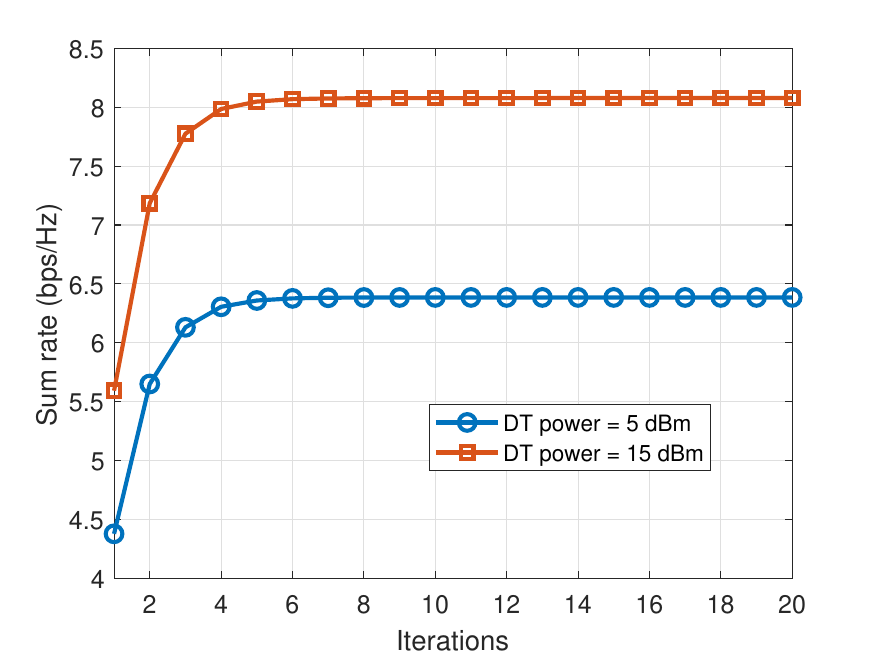}
\caption{Number of iterations versus sum rate of D2D communications for different DT power.}
\label{Fig3}
\end{figure}
%%%%%%%%%%%%%
%%%%%%%%%%%%%%%%
\begin{figure}[!t]
\centering
\includegraphics [width=0.48\textwidth]{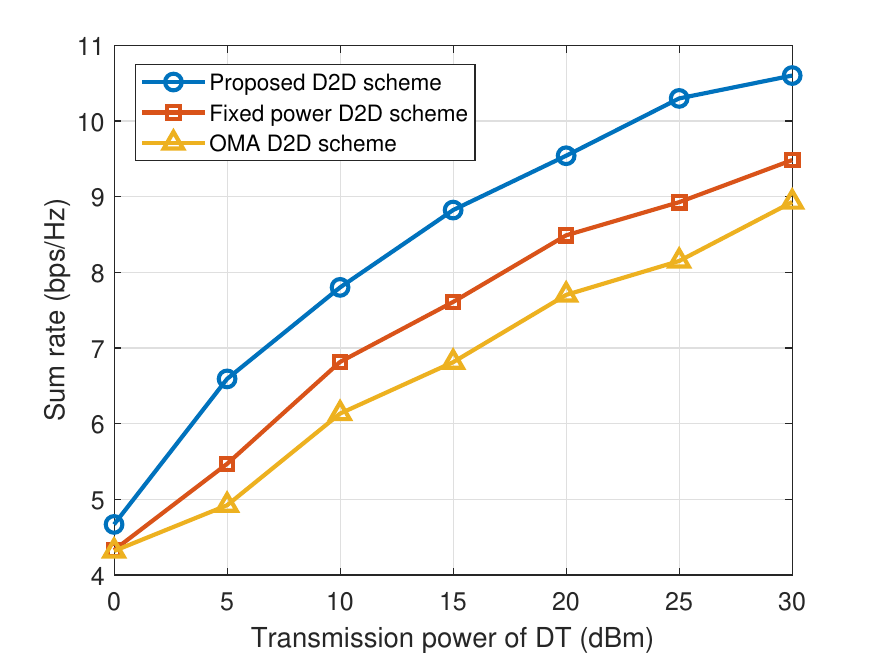}
\caption{Available transmit power of DT versus sum rate of D2D communications.}
\label{Fig2}
\end{figure}
%%%%%%%%%%%%%
%%%%%%%%%%%%%%%%
\begin{figure}[!t]
\centering
\includegraphics [width=0.48\textwidth]{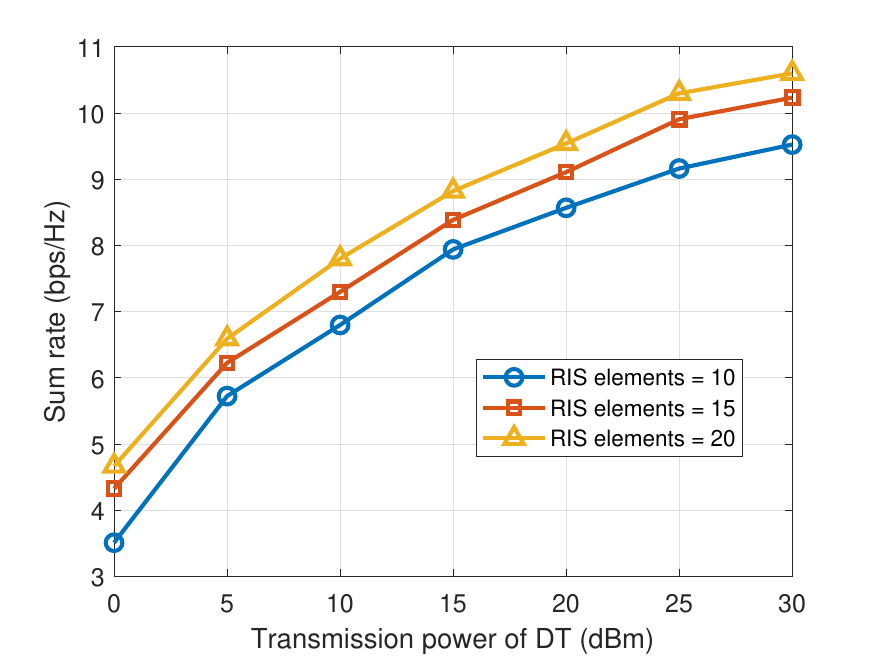}
\caption{Available transmit power of DT versus sum rate of D2D communications for different elements of RIS.}
\label{Fig1}
\end{figure}
%%%%%%%%%%%%%
\section{Results and Discussion}
We provide the simulation results in this section based on Monte Carlo simulations. Unless mentioned otherwise, the system parameters are set as the number of RIS elements as 20, the transmit power of DT and UAV is $P_t=30$ dBm, the minimum SINR is $K=20$ dB, the minimum SINR of cellular user is $\gamma_c=20$ dB, and noise variance is $\sigma^2=174$ dBm. For comparison, we also consider two benchmark schemes, i.e., the OMA D2D scheme with optimal power allocation at DT and optimal passive beamforming at RIS, while fixed power D2D scheme with fixed NOMA power allocation at DT and optimal passive beamforming at RIS. First, we analyze the convergence behaviour of the proposed scheme where the achievable sum rate of D2D communications is plotted against the number of iterations for two different DT transmit power (5 dBm and 15 dBm), as shown in Fig. \ref{Fig3}. We can see that the proposed D2D communications scheme converges within a few iterations, showing its low complexity. 

Next, we discuss the performance of the considered system for all three schemes by plotting the achievable sum rate against the available transmit power of DT, as illusterated in Fig. \ref{Fig2}. As can be seen, the achievable sum rate of all three schemes increases with the increasing transmit power of DT. However, the proposed D2D communications scheme achieves a high sum rate compared to the fixed power D2D and OMA D2D communications schemes. For instance, when the transmit power of DT is set as $P_t=25$ dBm, the achievable sum rate of the proposed D2D communications scheme is 10.3 bps/Hz, the fixed D2D communications scheme is 8.9 bps/Hz, and the OMA D2D communications scheme is 8.1 bps/Hz, respectively. In other words, the proposed D2D communications scheme achieves a 15\% higher sum rate compared to the fixed D2D communications scheme while a 27\% higher sum rate compared to the OMA D2D communications scheme.

To further analyze the performance of the proposed D2D communications scheme, it is important to demonstrate the impact of IRS elements. Fig \ref{Fig1} plots the achievable sum rate of the proposed D2D communications scheme against the available transmit power of DT considering different numbers of RIS elements. In this figure, we set the RIS elements as 10, 15, and 20. We can observe that the system with a large number of RIS elements achieves a high sum rate compared to the system with a small number of RIS elements. More specifically, for equal system parameters when the transmit power of DT is set as 25 dBm, the achievable sum rate of the proposed D2D communications with 20 RIS elements is 10.3 bps/Hz while 9.9 bps/Hz with 15 RIS elements and 9.1 bps/Hz with 10 RIS elements, respectively. It shows that RIS with large elements achieves high received signal gain than RIS with small elements.
\section{Conclusions}
The combination of D2D communications with NOMA and RIS can overcome spectrum and interference issues, improving the performance of wireless communications. This paper has considered a sum rate maximization problem for RIS enhanced NOMA D2D communications underlaying UAV networks. Our optimization scheme has optimized the power budget of the D2D transmitter, NOMA power allocation for D2D receivers and passive beamforming of RIS while ensuring the minimum SINR of the cellular user. The problem has been solved by adopting SCA and alternating optimization methods. Numerical results show that the proposed solution converges within a few iterations and achieves a higher sum rate than the benchmark solutions.

\bibliographystyle{IEEEtran}
\bibliography{Wali_EE}

\end{document}